%% file: eprint.tex
\documentclass[12pt]{article}
\usepackage{graphicx}

\newcommand\pubnumber{}
\newcommand\pubdate{\today}

\def\institute{Institut f\"ur Experimentelle Kernphysik\\
Department of Physics\\
Karlsruhe Instute of Technology, Karlsruhe, GERMANY}

\def\Title#1{\begin{center} {\Large #1 } \end{center}}
\def\Author#1{\begin{center}{ \sc #1} \end{center}}
\def\Address#1{\begin{center}{ \it #1} \end{center}}

\newcommand\pubblock{\rightline{\begin{tabular}{l} \pubnumber\\
         \pubdate  \end{tabular}}}
\newenvironment{Abstract}{\begin{quotation}  }{\end{quotation}}
\newenvironment{Presented}{\begin{quotation} \begin{center} 
             PRESENTED AT\end{center}\bigskip 
      \begin{center}\begin{large}}{\end{large}\end{center} \end{quotation}}

\input econfmacros.tex

\begin{document}
\begin{titlepage}
\pubblock

\vfill
\Title{Reconstruction and Selection of Single Top + Higgs to $\mathrm{b}\overline{\mathrm{b}}$ events at CMS in pp Collisions at $\sqrt{s}$ = 13 TeV}
\vfill
\Author{Kevin Fl\"oh\\
on behalf of the CMS Collaboration}
\Address{\institute}
\vfill
\begin{Abstract}
The associated production of a single top quark together with a Higgs boson at the L\textbf{•}HC can be used to lift the degeneracy regarding the sign of the top quark Yukawa coupling. Therefore, t-channel and tW-channel production where the Higgs boson is decaying into a $\mathrm{b}\overline{\mathrm{b}}$ pair is inspected. Boosted decision trees are used to assign jets to the quarks in the final state and classify the events. To maximize the separation between the signal and the dominating $\mathrm{t}\overline{\mathrm{t}}$ background each event is reconstructed under a signal and a $\mathrm{t}\overline{\mathrm{t}}$ hypothesis. The poster focuses on the event selection and the reconstruction of events.
\end{Abstract}
\vfill
\begin{Presented}
$9^{th}$ International Workshop on Top Quark Physics\\
Olomouc, Czech Republic,  September 19--23, 2016
\end{Presented}
\vfill
\end{titlepage}
\def\thefootnote{\fnsymbol{footnote}}
\setcounter{footnote}{0}

\section{Introduction}
\begin{figure}[htb]
\centering
\includegraphics[height=2in]{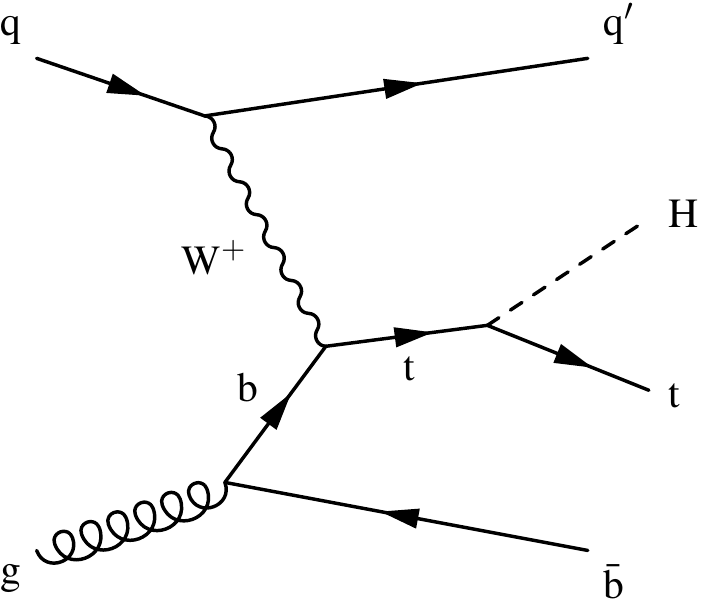}
\includegraphics[height=2in]{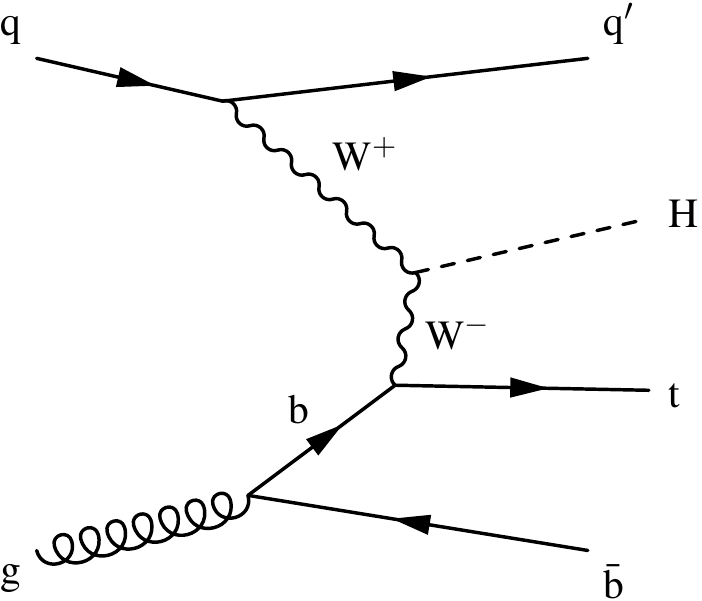}
\caption{The Higgs boson couples either to the W boson or to the top quark.}
\label{fig:Feynman}
\end{figure}

The associated production of a single top quark with a Higgs boson at the LHC can be used to lift the degeneracy regarding the sign of the top quark Yukawa coupling. In this analysis \cite{CMS:2016ygt} t-channel (tHq) and tW-channel (tHW) production is inspected where the Higgs boson decays to a  $\mathrm{b}\overline{\mathrm{b}}$ pair. The data taken in 2015 by the CMS detector \cite{Chatrchyan:2008aa}, corresponding to an integrated luminosity of $2.3 \mathrm{fb}^{-1}$ is analyzed.\\
As shown in Figure \ref{fig:Feynman} the Higgs boson couples either to the top quark (coupling parameter $\kappa_\mathrm{t}$) or to the W boson (coupling parameter $\kappa_\mathrm{V}$). The amplitude of this process is $A\propto (\kappa_\mathrm{V}-\kappa_\mathrm{t})$ in first approximation because of the interference of the two amplitudes. Hence, the cross section and the kinematic variables are sensitive to the strength and sign of the Higgs couplings. 51 points in the $\kappa_\mathrm{V}$-$\kappa_\mathrm{t}$-plane are examined with $-3\leq\kappa_\mathrm{t}\leq3$ and $0.5\leq\kappa_\mathrm{V}\leq1.5$.\\
The standard model (SM) predicts $\kappa_\mathrm{t}=\kappa_\mathrm{V}=1$, whereas the inverted top coupling scenario (ITC), which is the most interesting non standard model point, has $\kappa_\mathrm{t}=-1$ and $\kappa_\mathrm{V}=1$. The cross section for the tHq process is 71 fb  and 16 fb for the tHW process in the SM, while it is enhanced by a factor of about eleven to 739 fb and 147 fb respectively for the ITC scenario.

\section{Event Selection}
\begin{figure}[htb]
\centering
\includegraphics[height=2in]{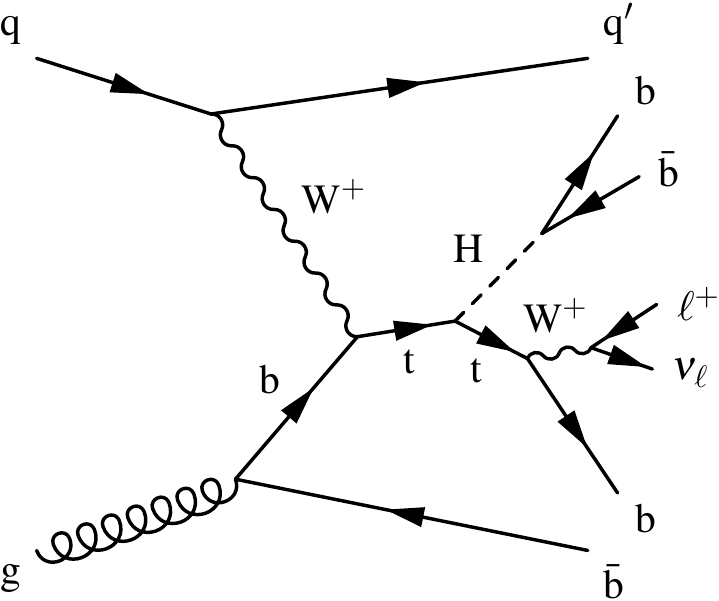}
\caption{Decay diagram of the leptonically decaying tHq event}
\label{fig:decay}
\end{figure}

Several requirements must be fulfilled to get a signal-enriched phase space. As shown in Figure \ref{fig:decay} only events with a leptonically decaying W boson are considered. Therefore, exactly one lepton with $p_\mathrm{T}>30 (25)\, \mathrm{GeV}$, $|\eta|<2.1 (2.4)$ in the e ($\mu$) + jets channel is required. Additionally a missing transverse energy of at least 45 (35) GeV is required.\\
The control region requires at least four jets with $p_\mathrm{T}>30\, \mathrm{GeV}$ and exactly two of them with a b tag. The 3 tag signal region has at least four jets with $p_\mathrm{T}>30\, \mathrm{GeV}$ with exactly three of them with a b tag. In the 4 tag signal region at least five jets with $p_\mathrm{T}>30\, \mathrm{GeV}$ and exactly four of them with a b tag are required. The control region and the two signal regions are completely dominated by semileptonic $\mathrm{t}\overline{\mathrm{t}}$ background events.

\section{Event Reconstruction}
\begin{figure}[htb]
\centering
\includegraphics[width=0.45\textwidth]{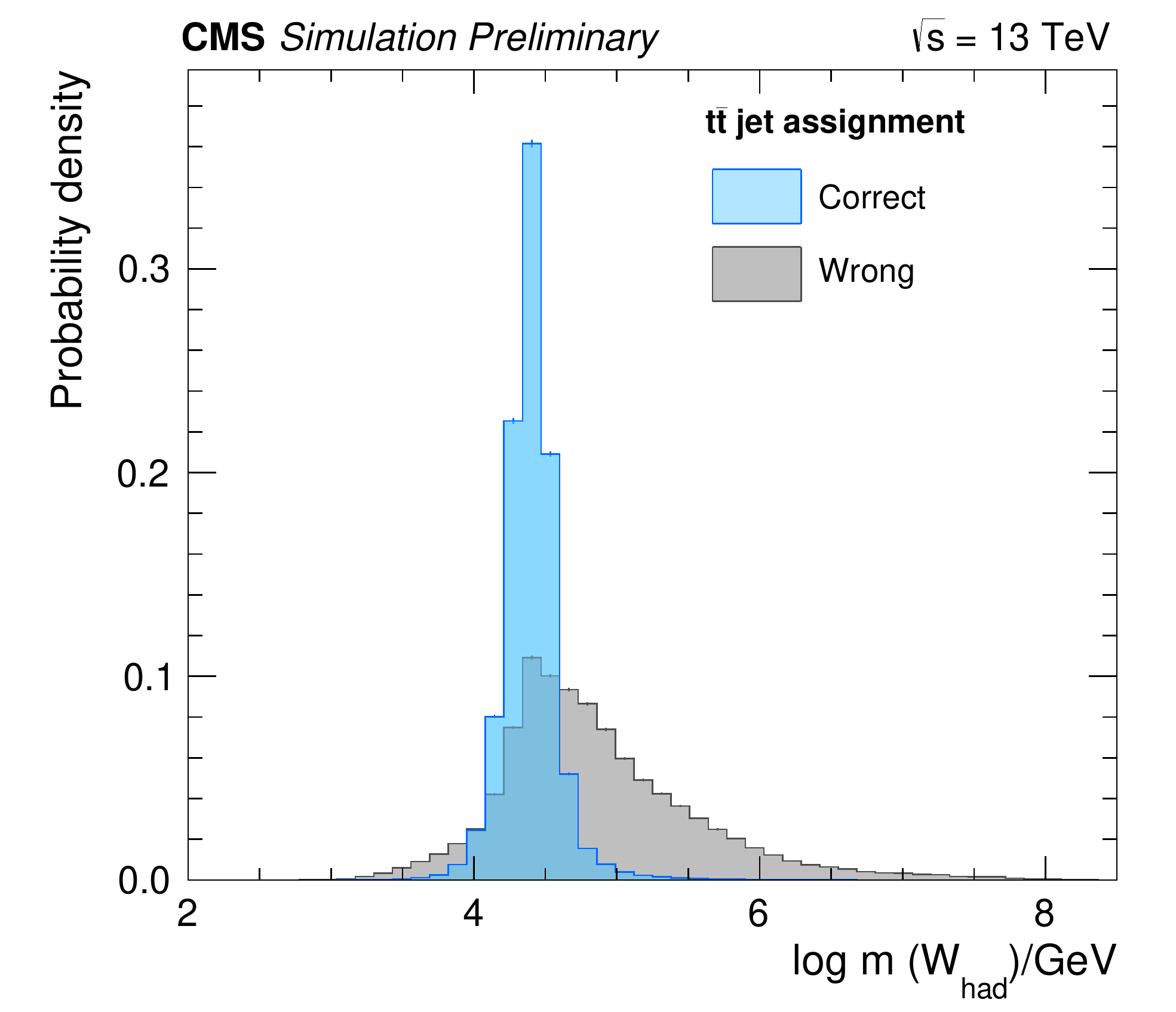}
\includegraphics[width=0.45\textwidth]{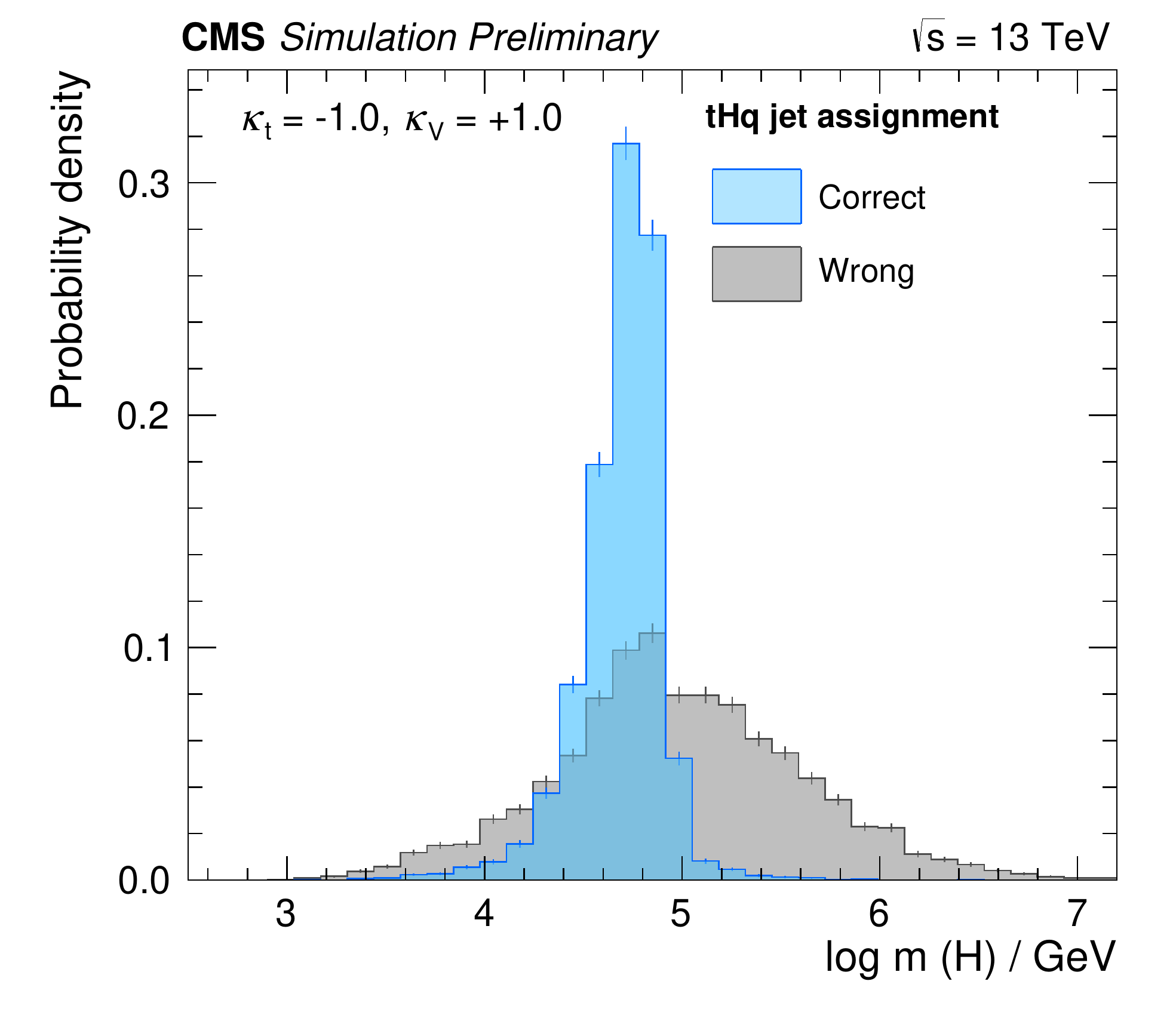}
\caption{Most important variables for the reconstruction: the mass of hadronically decaying W boson (left, $\mathrm{t}\overline{\mathrm{t}}$ reconstruction) and the mass of the Higgs boson (right, signal reconstruction). Taken from \cite{CMS:2016ygt}.}
\label{fig:Input}
\end{figure}
Boosted decision trees (BDTs) are used to assign jets to the quarks in the final state. To maximize the separation between the signal and the dominating $\mathrm{t}\overline{\mathrm{t}}$ background each event is reconstructed both under the assumption of being a signal event and under the assumption of being a semileptonic $\mathrm{t}\overline{\mathrm{t}}$ event. Therefore, the BDTs are trained to separate between the correct and a random wrong jet assignment.\\
For the $\mathrm{t}\overline{\mathrm{t}}$ hypothesis one BDT is trained while for the signal hypotheses 51 BDTs have to be trained (one for each point in the $\kappa_\mathrm{V}$-$\kappa_\mathrm{t}$-plane). The invariant masses and the pseudorapidity of the top quark and the Higgs boson as well as the b tag discriminants of the b quarks are used as input variables. The most important variables for each reconstruction are shown in Figure \ref{fig:Input}.\\
In the application of the BDTs, all possible jet permutations are considered, the BDT score is calculated for each permutation and the assignment with the highest BDT score is picked. This is done both under the signal assumption and the $\mathrm{t}\overline{\mathrm{t}}$ assumption. In the $\mathrm{t}\overline{\mathrm{t}}$ reconstruction jets matched to bottom quarks are required to be b-tagged. The tHq reconstruction requires the jet matched to the recoil jet to be untagged and the jets matched to bottom quarks are required to be central.

\begin{figure}[htb]
\centering
\includegraphics[width=0.45\textwidth]{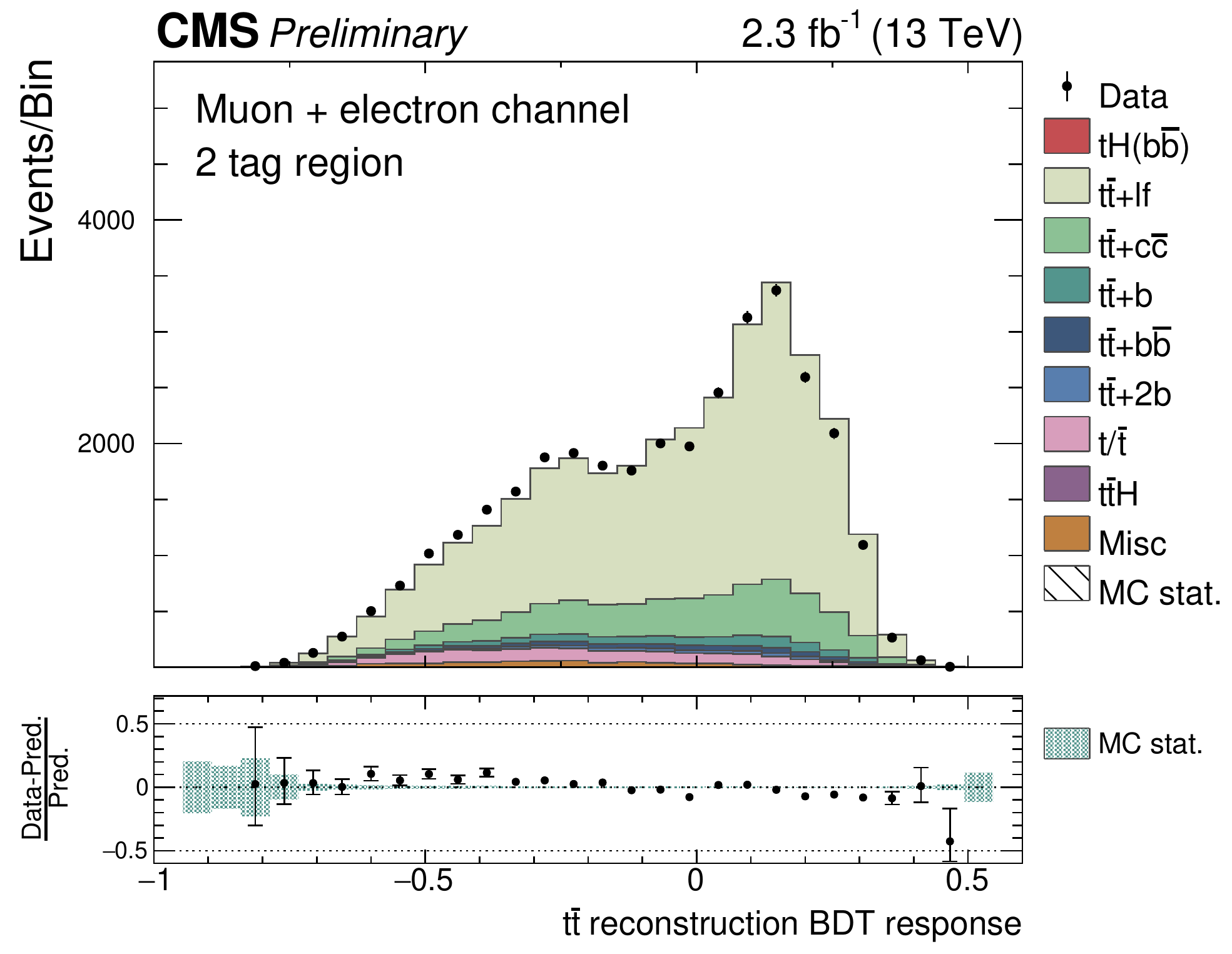}
\includegraphics[width=0.45\textwidth]{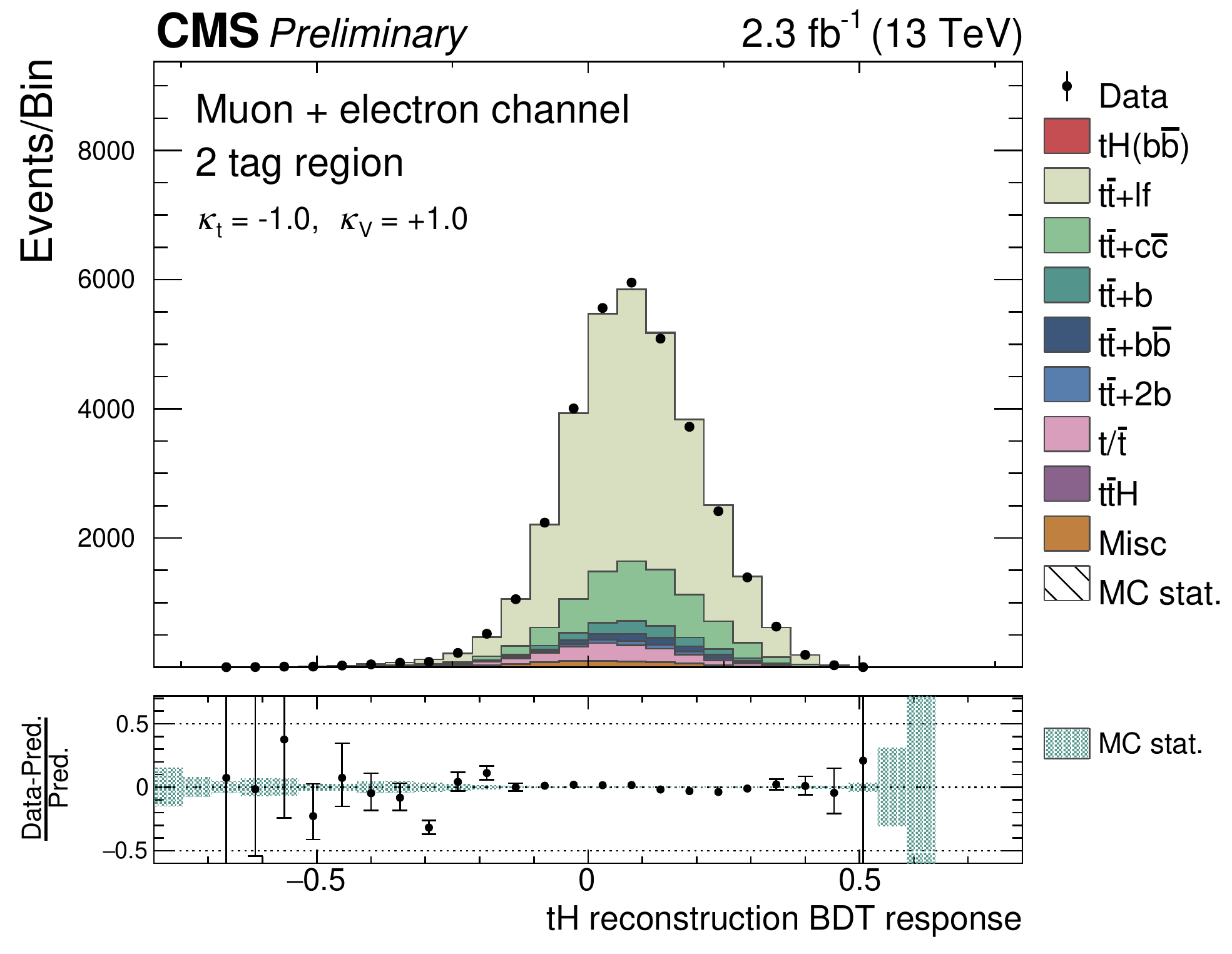}
\caption{BDT output of $\mathrm{t}\overline{\mathrm{t}}$ (left) and signal hypothesis (right). Taken from \cite{CMS:2016ygt}.}
\label{fig:dataMC}
\end{figure}

\begin{figure}[htb]
\centering
\includegraphics[width=0.45\textwidth]{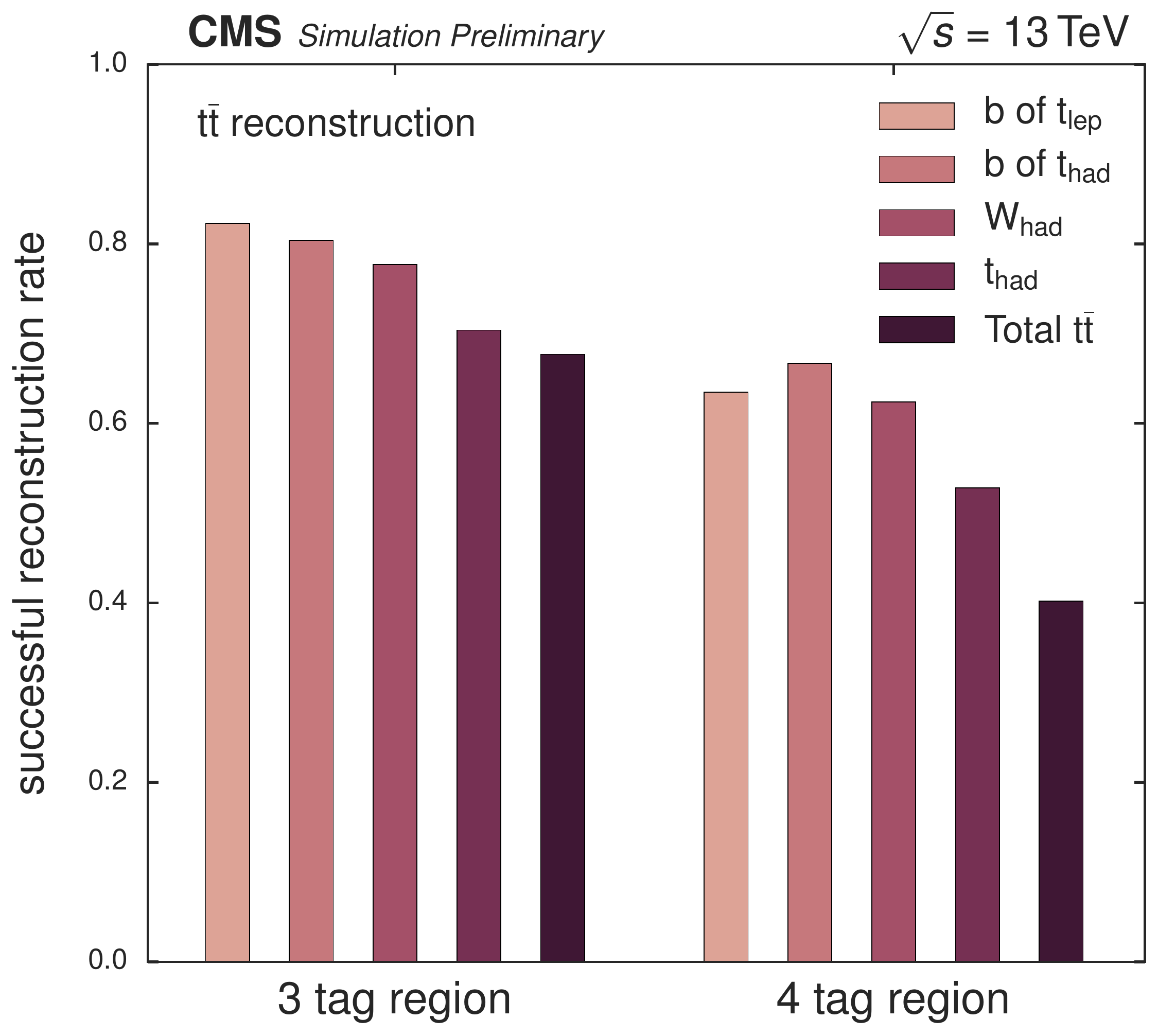}
\includegraphics[width=0.45\textwidth]{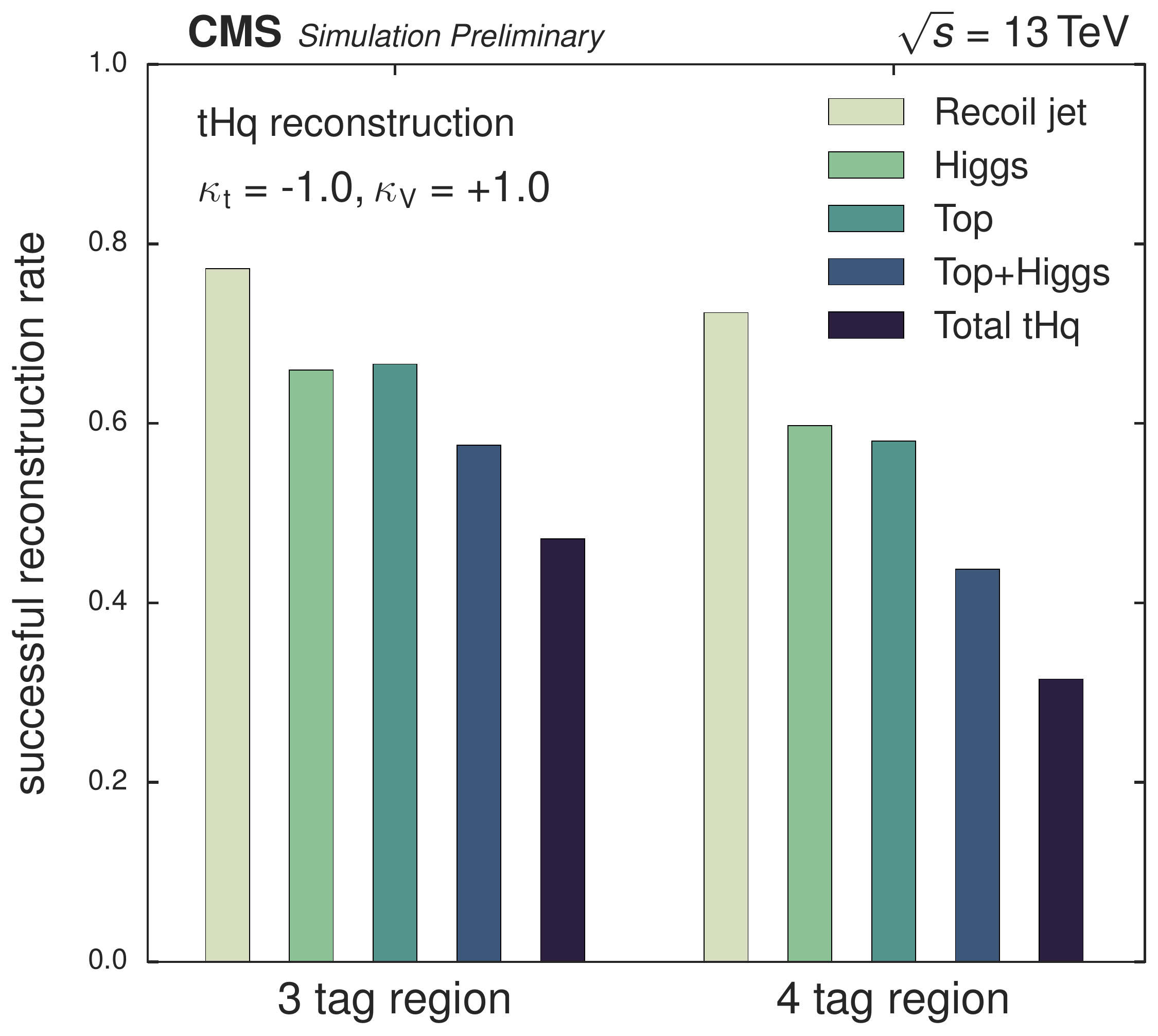}
\caption{Successful reconstruction rate for $\mathrm{t}\overline{\mathrm{t}}$ (left) and signal hypothesis (right). Taken from \cite{CMS:2016ygt}.}
\label{fig:reco}
\end{figure}

The reconstruction results in specific variables as the Higgs boson mass for the signal hypothesis and as the mass of the hadronically decaying W boson mass for the background hypothesis. For a full list of the variables see \cite{CMS:2016ygt}.\\
The agreement between data and simulation for the BDT response is shown in Figure \ref{fig:dataMC}. The rate of reconstructing a whole $\mathrm{t}\overline{\mathrm{t}}$ event correctly is 68\% (40\%)  and for whole a tHq event 47\% (32\%) in the 3 tag (4 tag) region, see Figure \ref{fig:reco}. 

\section{Results}
The variables which are generated by reconstructing the processes and additionally, some global variables are used in 51 final classification BDTs (one for each point) to separate between signal and background processes. Finally, a fit on the classification BDT response provides an exclusion limit for each point in the $\kappa_\mathrm{V}$-$\kappa_\mathrm{t}$-plane.\\
The observed (expected) upper limits on the combined production rates of tHq and tHW processes are $113.7~(98.6) \times \sigma_\mathrm{SM}$ for the SM scenario and $6.0~(6.4) \times \sigma_\mathrm{ITC}$.\\
For further details on the classification, limit setting and the exclusion limits on the other points see \cite{CMS:2016ygt}.

\end{document}

%% file: econfmacros.tex
\def\beq{\begin{equation}}
\def\eeq#1{\label{#1}\end{equation}}
\def\eeqn{\end{equation}}

\def\beqa{\begin{eqnarray}}
\def\eeqa#1{\label{#1}\end{eqnarray}}
\def\eeqan{\end{eqnarray}}

\let\bar=\overbar

\def\Dslash{\not{\hbox{\kern-4pt $D$}}}
\def\dslash{\not{\hbox{\kern-2pt $\del$}}}

\def\msb{{\bar{\ssstyle M \kern -1pt S}}}

